\documentclass{article}

\usepackage{amssymb,nath}

\def\eqref#1{{\rm(\ref{#1})}}
\def\Cal#1{{{\cal #1}}}

\let\frak\mathfrak
\begin{document}

\title{Recursion operators for vacuum Einstein equations with  symmetries} 
\author{M. Marvan 
\\\footnotesize Mathematical Institute, Silesian University in Opava, 
\\\footnotesize Na Rybn\'\i\v cku 1, 746 01 Czech Republic}
\date{}
\maketitle

\abstract{Direct and inverse recursion operator is derived for the vacuum 
Einstein equations for metrics with two commuting Killing vectors that
are orthogonal to a foliation by 2-dimensional leaves.}

\section {Introduction}

In the past decade, inverse recursion operators became a subject of a number
of papers~\cite{G,G-H,L,L-C}.
In particular, the work of Guthrie~\cite{G} opened a new perspective on 
recursion operators by essentially identifying them with auto-B\"acklund 
transformations for linearized equation~\cite{M2}.

It is known for a long time that recursion operators of integrable systems 
are obtainable from their Lax pairs (see \cite{G-K-S} and references therein)
and ZCR's~(see~\cite{K-K-Sa,Sa}).
However, recently it became clear that zero curvature representations are 
related much closer to inverse recursion operators than to their `direct' 
counterparts~\cite{M5,M-S}.
Examples that have been already published elsewhere include the 
Korteweg--de Vries and Tzitz\'eica equation~\cite{M5} and the 
stationary Nizhnik--Novikov--Veselov equation~\cite{M-S}.
In the present paper, the methods of~\cite{M5,M-S} are applied to equations
of General Relativity.

\section{Recursion operators}

Let $\Cal E = \{F^l = 0\}$ be a system of PDE's in  unknown functions $u^k$ 
of two independent variables $x,y$.
We assume that $F^l$ are functions of $x,y,u^k$ and a finite number of the 
derivatives 
\smash{$u^k_{ij} = \frac{\partial^{i+j} u^k}{\partial x^i \partial y^j}$}
(\smash{$u^k_{00} = u^k$}).
Consider the infinite-dimensional jet space $J^\infty$ 
with local coordinates $x,y,u^k_{ij}$ along with the commuting vector fields 
\smash{$D_x = \frac \partial{\partial x}
 + \sum_{ij} u^k_{i+1,j} \frac \partial{\partial u^k_{ij}}$},
\smash{$D_y = \frac \partial{\partial y}
 + \sum_{ij} u^k_{i,j+1} \frac \partial{\partial u^k_{ij}},$}
called {\it total derivatives}.
The submanifold $E$ determined by equations $F^l = 0$ and their
differential consequences
$D_x F^l = 0$, $D_y F^l = 0$, $D_x^2 F^l = 0$, $D_x D_y F^l = 0$, 
$D_y^2 F^l = 0$, \dots, is called the {\it equation manifold} 
(and is an underlying space of the diffiety structure~\cite{B-V-V} 
employed in \cite{M5}).
In this context, infinitesimal symmetries (more precisely, their
generating functions) are functions $U^k$ defined on $E$ such
that 
$\sum_{k,i,j} \frac{\partial F^l}{\partial u^k_{ij}} D_x^i D_y^j U^k = 0.$

It is then natural to consider the jet space with coordinates 
$x,y,\ u^k_{ij},\ U^k_{ij}$, and denote
\begin{equation}
\label{L}
LF^l = \sum_{k,i,j} \frac{\partial F^l}{\partial u^k_{ij}} U^k_{ij}.
\end{equation}
The system $L \Cal E := \{F^l = 0, LF^l = 0\}$ on unknowns $u^k,U^k$ 
will be called the {\it linearized equation}.
Now, Guthrie's recursion operators~\cite{G} may be interpreted as 
auto-B\"acklund transformations of the linearized equation $L \Cal E$ that 
keep variables $u^k$ unchanged~\cite{M5}.

In now standard formalism~\cite{O}, recursion operators are 
pseudodifferential operators, characterized by the occurrence of 
inverses of total derivatives~$D_x^{-1}$. 
Under Guthrie's approach, $p = D_x^{-1} f$ is introduced as an 
auxiliary nonlocal variable satisfying
\begin{equation}
\label{p}
p_x = f, \qquad 
p_y = g,
\end{equation}
provided such a $g$ exists, and it actually does without known exception;
see Sergyeyev~\cite{Se} for a proof in case of evolution systems.
Thus, $p$ is a potential of a conservation law $f\,dx + g\,dy$ of the 
linearized equation $L \Cal E$.

For the inverse recursion operators, the nonlocalities tend to be 
genuinely nonabelian pseudopotentials related to a zero curvature 
representation of the system in question.
Let $\frak g$ be a matrix Lie algebra.
Let $\alpha = A\,dx + B\,dy$ be a $\frak g$-valued zero curvature 
representation (ZCR) for the system $\Cal E$.
This means that $A,B$ are $\frak g$-valued functions on the equation 
submanifold $E$ and $D_y A - D_x B + [A,B] = 0$ holds on~$E$.
Let us introduce the associated pseudopotential $P$ as a $\frak g$-valued
solution of the compatible system
\begin{equation}
\label{P}
P_x = [A,P] + LA, \qquad
P_y = [B,P] + LB.
\end{equation}
A recursion operator $R$ is then a linear operator in $U^k$ and $P$
such that $U' = R(U,P)$ solves the linearized system $L \Cal E$ whenever 
$U$ does and $P$ satisfies~(\ref{P}) (see~\cite{M5,M-S}).
In this way, the inverse recursion operator can be found without 
previous knowledge of the direct recursion operator.
A remarkable aspect of this approach is that $R(U,P)$ tends to be a very 
simple expression. 

For the above scheme to work, it is not necessary that the ZCR $\alpha$
a priori depends on the ``spectral parameter.''
However, if $R$ is a recursion operator related to the ZCR $\alpha$ as
above, then $(R^{-1} + \mu`Id)^{-1}$ is another recursion operator, 
associated with a ZCR $\alpha_\mu$ which depends on $\mu$.

\section{The results}

We consider vacuum Einstein equations for a space-time with two commuting
Killing vectors that are orthogonal to a foliation by 2-dimensional 
surfaces~\cite{Ge1,Ge2}.
Our presentation will be restricted to the case when both Killing vectors 
are space-like.
The case when one of the Killing vectors is time-like is equivalent to 
ours via an appropriate complex transformation of coordinates.

As is well known, there exist coordinates $x,y,z^1,z^2$ such that the 
metric in question can be written in the form 
$ds^2 = 2 f(x,y)\,dx\,dy + g_{ij}(x,y)\,dz^i\,dz^j$ 
(the Lewis~\cite{Le} metric).
The vacuum Einstein equations essentially reduce to
\begin{equation}
\label{g}
(\sqrt{`det g}\, g_x g^{-1})_y + (\sqrt{`det g}\, g_y g^{-1})_x = 0,
\end{equation}
while $f$ can be obtained by quadrature.
Using the standard normalization $`det g = (x + y)^2$ compatible with 
Eq.~(\ref{g}), we parametrize $g$ as follows:
$g_{11} = (x + y)/u$, 
$g_{12} = (x + y)v/u$, 
$g_{22} = (x + y)(u^2 + v^2)/u$. 
Equation (\ref{g}) then becomes
\begin{equation}
\label{uv}
u_{xy} = \frac{u_x u_y - v_x v_y}{u} - \frac{1}{2} \frac{u_x + u_y}{x + y},
\\
v_{xy} = \frac{v_x u_y + u_x v_y}{u} - \frac{1}{2} \frac{v_x + v_y}{x + y}.
\end{equation}
As is well known, Eq.~(\ref{uv}) has a ZCR and a B\"acklund 
transformation~\cite{B-Z,Ch,D-K-M,H,Ma1,Ma2,O-W}.
The ZCR reads
$$
A = \frac{1}{2} \left(\begin{array}{rc} 
 -(\theta + 1) {u_x}/{u} & 
 (\theta + 1) {v_x}/{u^2} \\ 
 (\theta - 1) v_x & 
 (\theta + 1) {u_x}/{u} 
\end{array}\right),
\\
B = \frac{1}{2 \theta}
 \left(\begin{array}{cc} 
 -(\theta + 1) {u_y}/{u} & 
 (\theta + 1) {v_y}/{u^2} \\ 
 (-\theta + 1) {v_y} & 
 (\theta + 1) {u_y}/u
 \end{array}\right),
$$
where
$\theta = \sqrt{\frac{\mu + y}{\mu - x}}$, $\mu$ being the spectral 
parameter.

The main result of this paper, obtained by the methods of~\cite{M5,M-S}, 
is as follows:
If nonlocal variables $p_{11},p_{12},p_{21}$ satisfy
\begin{equation}
\label{pij}
\padded{\qqquad}
p_{11,x} =  -\frac{\theta - 1}{2} v_x p_{12}  
 + \frac{\theta + 1}{2} \frac{v_x}{u^2} p_{21}
 - \frac{\theta + 1}{2} \frac1{u} U_x  
 + \frac{\theta + 1}{2} \frac{u_x}{u^2} U,
\\
p_{12,x} =  -(\theta + 1) \frac{v_x}{u^2} p_{11}
 - (\theta + 1) \frac{u_x}{u} p_{12} 
\\ - (\theta + 1) \frac{v_x}{u^3} U 
 + \frac{\theta + 1}{2} \frac1{u^2} V_x,
\\
p_{21,x} = (\theta - 1) v_x p_{11}  
 + (\theta + 1) \frac{u_x}{u} p_{21}
 + \frac{\theta - 1}{2} V_x,
\\
p_{11,y} =  \frac{\theta - 1}{2 \theta} v_y p_{12}  
 + \frac{\theta + 1}{2 \theta} \frac{v_y}{u^2} p_{21}  
 + \frac{\theta + 1}{2 \theta} \frac{u_y}{u^2} U 
 - \frac{\theta + 1}{2 \theta} \frac1{u} U_y,
\\
p_{12,y} =  -\frac{\theta + 1}{\theta} \frac{v_y}{u^2} p_{11}  
 - \frac{\theta + 1}{\theta} \frac{u_y}{u} p_{12}  
\\ - \frac{\theta + 1}{\theta} \frac{v_y}{u^3} U 
 + \frac{\theta + 1}{2 \theta} \frac1{u^2} V_y,
\\ 
p_{21,y} =  -\frac{\theta - 1}{\theta} v_y p_{11}  
 + \frac{\theta + 1}{\theta} \frac{u_y}{u} p_{21} 
 - \frac{\theta - 1}{2 \theta} V_y,
\return
$$
then
\begin{equation}
\label{iro}
U' = 2 \frac{u}{\sqrt{(\mu - x)(\mu + y)}} p_{11}
 + \frac{1}{\sqrt{(\mu - x)(\mu + y)}} U,
\\
V' = -\frac{u^2}{\sqrt{(\mu - x)(\mu + y)}} p_{12} 
  - \frac{1}{\sqrt{(\mu - x)(\mu + y)}} p_{21}
\end{equation}
is a recursion operator for Eq. (\ref{uv}), namely, it sends symmetries
to symmetries if the latter are viewed as solutions of the
linearized system 
$$
\padded{\qquad}
U_{xy} = (\frac{u_y}{u} - \frac{1}{2(x + y)}) U_x
 + (\frac{u_x}{u} - \frac{1}{2(x + y)}) U_y
\\ - \frac{u_x u_y - v_x v_y}{u^2} U
 - \frac{v_y}{u} V_x
 - \frac{v_x}{u} V_y,
\\
V_{xy} = \frac{v_y}{u} U_x 
 + \frac{v_x}{u} U_y 
 - \frac{v_x u_y + u_x v_y}{u^2} U
\\ + (\frac{u_y}{u} - \frac{1}{2(x + y)}) V_x
 + (\frac{u_x}{u} - \frac{1}{2(x + y)}) V_y.
\return
$$

The `direct' recursion operator for this equation seems to be missing in the 
literature; we can obtain it by inverting the operator~(\ref{iro}), the
result being
$$
U' = u v p_1 - u p_2 + (y - x) U,
\\
V' = - \frac{1}{2} (u^2 - v^2) p_1
 - v p_2
 - \frac{1}{2} p_3
 + (y - x) V, 
$$
where $p_1,p_2,p_3$ satisfy
$$
p_{1,x} = (x + y) (-2 \frac{v_x}{u^3} U + \frac{1}{u^2} V_x), 
\\
p_{2,x} = (x + y) (-\frac{u u_x + 2 v v_x}{u^3} U
 + \frac{1}{u} U_x
 + \frac{v_x}{u^2} V
 + \frac{v}{u^2} V_x),
\\
p_{3,x} = (x + y) (\wall 2 \frac{(u u_x + v v_x) v}{u^3} U
 - 2 \frac{v}{u} U_x
\\ - 2 \frac{u u_x + v v_x}{u^2} V
 + \frac{u^2 - v^2}{u^2} V_x),
\return
\\
p_{1,y} = (x + y) (2 \frac{v_y}{u^3} U - \frac{1}{u^2} V_y), 
\\
p_{2,y} = (x + y) (\frac{u u_y + 2 v v_y}{u^3} U
 - \frac{1}{u} U_y
 - \frac{v_y}{u^2} V 
 - \frac{v}{u^2} V_y),
\\
p_{3,y} = (x + y) (\wall -2 \frac{(u u_y + v v_y) v}{u^3} U
 + 2 \frac{v}{u} U_y
\\ + 2 \frac{u u_y + v v_y}{u^2} V
 - \frac{u^2 - v^2}{u^2} V_y).
\return
$$
It is readily seen that $p_i$ are potentials of the linearizations ~\cite{M2} 
of the three obvious conservation laws of Eq.~(\ref{g}).

Quite unusually, neither of the recursion operators found generates an 
infinite series of local symmetries (and no such series is known).
The action of our operators on the infinite-dimensional Geroch group of 
nonlocal symmetries~\cite{Ge2,Ki} remains to be investigated.

It is convenient to rewrite system~(\ref{pij}) in triangular form.
To achieve this, we introduce the Riccati pseudopotential $q$ by
$$
q_x = \frac{\theta - 1}{2} v_x q^2
 - (\theta + 1) \frac{u_x}{u} q
 - \frac{\theta + 1}{2} \frac{v_x}{u^2},
\\
q_y =  -\frac{\theta - 1}{2\theta} v_y q^2
 - \frac{\theta + 1}{2\theta} \frac{u_y}{u} q 
 - \frac{\theta + 1}{2\theta} \frac{v_y}{u^2}
$$
and a nonlocal potential $r$ by
$$
r_x = (\theta - 1) v_x q - (\theta + 1) \frac{u_x}{u},
\\
r_y = -\frac{\theta - 1}{\theta} v_y q
 - \frac{\theta + 1}{\theta} \frac{u_y}{u} .
$$
Then the inverse recursion operator assumes the form
$$
U' = \frac 1{\sqrt{(\mu - x)(\mu + y)}}
 (2 u Q - 2 \frac{u q}{\mathrm e^{r}} R + U),
\\
V' = \frac 1{\sqrt{(\mu - x)(\mu + y)}}
(-u^2 \mathrm e^{r} P - 2 u^2 q Q
 + \frac{u^2 q^2 - 1}{\mathrm e^{r}} R),
$$
where $P,Q,R$ are supposed to satisfy
$$
\padded{\qquad}
P_x = (\theta + 1) \frac{q}{u} \mathrm e^{-r} U_x
 + (\frac{\theta + 1}{2} \frac1{u^2} - \frac{\theta - 1}{2} q^2)
 \mathrm e^{-r} V_x
\\ - (\theta + 1) (\frac{q u_x}{u^2} + \frac{v_x}{u^3}) \mathrm e^{-r} U,
\\
Q_x =  - \frac{\theta + 1}{2} \frac{1}{u} U_x
 + \frac{\theta - 1}{2} q V_x 
 + \frac{\theta + 1}{2} \frac{u_x}{u^2} U
 - \frac{\theta - 1}{2} v_x \mathrm e^{r} P,
\\
R_x = \frac{\theta - 1}{2} \mathrm e^{r} V_x
 + (\theta - 1) v_x \mathrm e^{r} Q,
\\
P_y = \frac{\theta + 1}{\theta} \frac{q}{u} \mathrm e^{-r} U_y
 + (\frac{\theta + 1}{2\theta} \frac1{u^2} + \frac{\theta - 1}{2\theta} q^2)
 \mathrm e^{-r} V_y
\\ - \frac{\theta + 1}{\theta} (\frac{q u_y}{u^2} + \frac{v_y}{u^3})
 \mathrm e^{-r} U,
\\
Q_y =  - \frac{\theta + 1}{2\theta} \frac1{u} U_y
 - \frac{\theta - 1}{2\theta} q V_y
 + \frac{\theta + 1}{2\theta} \frac{u_y}{u^2} U 
 + \frac{\theta - 1}{2\theta} v_y \mathrm e^{r}P,
\\
R_y =  - \frac{\theta - 1}{2\theta} \mathrm e^{r} V_y
 - \frac{\theta - 1}{\theta} v_y \mathrm e^{r} Q.
\return
$$
This form of the inverse recursion operator is better adapted to generation 
of symmetries, which is, however, beyond the scope of this paper.

\section*{Acknowledgements}

I would like to thank G. Vilasi for drawing my attention to the problem.
The support from the grant MSM:J10/98:192400002 is gratefully acknowledged.

\end{document}